\documentclass[12pt]{spieman}  % 12pt font required by SPIE;
\usepackage{amsmath,amsfonts,amssymb}
\usepackage[mathscr]{euscript}
\usepackage{graphicx}
\usepackage{setspace}
\usepackage{tocloft}
\usepackage{xcolor}
\usepackage{lineno}
\usepackage{soul}
\usepackage[super]{nth}

\title{The Three-Sided Pyramid Wavefront Sensor. I. Simulations and Analysis for Astronomical Adaptive Optics}

\author[a]{Lauren Schatz}
\author[b]{Jared R. Males}
\author[c]{Carlos Correia}
\author[c]{Benoit Neichel}
\author[c]{Vincent Chambouleyron}
\author[d]{Johanan Codona}
\author[e]{Olivier Fauvarque}
\author[c]{Jean-François Sauvage}
\author[f,c]{Thierry Fusco}
\author[d]{Michael Hart}
\author[c]{Pierre Janin-Potiron}
\author[g]{Robert Johnson}
\author[b]{Joseph Long}
\author[g]{Mala Mateen}

\affil[a]{University of Arizona, Wyant College of Optical Sciences, Tucson, Arizona, United States}
\affil[b]{University of Arizona, Steward Observatory, Tucson, Arizona, United States}
\affil[c]{Aix Marseille Univ, CNRS, CNES, LAM, Marseille, France}
\affil[d]{Hart Scientific Consulting International LLC, Tucson, Arizona, United States}
\affil[e]{IFREMER, Laboratoire Detection, Capteurs et Mesures (LDCM), Centre Bretagne, ZI de la Pointe du Diable, CS 10070, 29280, Plouzane, France}
\affil[f]{DOTA, ONERA, Université Paris Saclay, Palaiseau, France}
\affil[g]{Air Force Research Lab, Starfire Optical Range, Kirtland Air Force Base, New Mexico, United States}

\cftpagenumbersoff{figure}
\cftpagenumbersoff{table}

\begin{document} 
\maketitle

\begin{abstract}

The Giant Segmented Mirror Telescopes (GSMTs) including the Giant Magellan Telescope (GMT), the Thirty Meter Telescope (TMT), and the European Extremely Large Telescope (E-ELT), all have extreme adaptive optics (ExAO) instruments planned that will use pyramid wavefront sensors (PWFS). The ExAO instruments all have common features: a high-actuator-count deformable mirror running at extreme speeds (greater than 1 kHz); a high-performance wavefront sensor (WFS); and a high-contrast coronagraph. A current limitation in ExAO wavefront sensor performance is the detector because ExAO needs high spatial sampling of the wavefront which requires a large detector. For ExAO instruments for the next generation of telescopes, alternative architectures of WFS are under consideration because there is a trade-off between detector size, speed, and noise that reduces the performance of GSMT-ExAO wavefront control. One option under consideration for a GSMT-ExAO wavefront sensor is a three-sided PWFS (3PWFS). The 3PWFS creates three copies of the telescope pupil for wavefront sensing, compared to the conventional four-sided PWFS (4PWFS) which uses four pupils. The 3PWFS uses fewer detector pixels than the 4PWFS and should therefore be less sensitive to read noise. Here we develop a mathematical formalism based on the diffraction theory description of the Foucault knife edge test that predicts the intensity pattern after the PWFS. Our formalism allows us to calculate the intensity in the pupil images formed by the PWFS in the presence of phase errors corresponding to arbitrary Fourier modes. We use these results to motivate how we process signals from a 3PWFS. We compare the Raw Intensity method which uses the signal in the pupils as is and derive the Slopes Maps calculation for the 3PWFS which combines the three pupil images of the 3PWFS to obtain the X and Y slopes of the wavefront. We then use the Object Oriented MATLAB Adaptive Optics toolbox (OOMAO) to simulate an end-to-end model of an adaptive optics system using a PWFS with modulation and compare the performance of the 3PWFS to the 4PWFS. In the case of a low read noise detector, the Strehl ratios of the 3PWFS and 4PWFS are within 0.01. When we included higher read noise in the simulation, we found a Strehl ratio gain of 0.036 for the 3PWFS using Raw Intensity over the 4PWFS using Slopes Maps at a stellar magnitude of 10. At the same magnitude the 4PWFS RI also outperformed the 4PWFS SM, but the gain was only 0.012 Strehl. This is significant because 4PWFS using Slopes Maps is how the PWFS is conventionally used for AO wavefront sensing. We have found that the 3PWFS is a viable wavefront sensor that can fully reconstruct a wavefront and produce a stable closed-loop with correction comparable to that of a 4PWFS, with modestly better performance for high read-noise detectors.

\end{abstract}

% Include a list of up to six keywords after the abstract
\keywords{adaptive optics, wavefront sensing, instrumentation, high contrast imaging, simulation}

\begin{spacing}{2}   % use double spacing for rest of manuscript

\section{Introduction}
\label{sect:intro}  % \label{} allows reference to this section

Within the next decade, the world will see a new generation of Giant Segmented Mirror Telescopes (GSMTs). The Giant Magellan Telescope\cite{fanson2020overview} (GMT) under construction at Las Campanas Observatory, Chile, will have seven 8.4-meter mirrors, forming a 24.5-meter primary mirror. The Thirty Meter Telescope\cite{chisholm2020thirty} (TMT), and the European Extremely Large Telescope\cite{ramsay2020eso} (E-ELT) at Cerro Paranal, Chile, have highly segmented primary mirrors. The 39-meter E-ELT primary mirror will be comprised of hundreds of 1.4-meter hexagonal segments. The TMT will have 492\cite{sanders2013thirty} 1.44-meter hexagonal segments to form the 30-meter primary mirror. The GSMTs in combination with extreme adaptive optics (ExAO) systems will have the angular resolution and light-collecting power to image Earth-like planets around M-dwarf stars\cite{males2019gmagao}. 

There are several ExAO instruments in operation on current generation telescopes. Current instruments include the Subaru Coronagraphic Extreme Adaptive Optics instrument (SCExAO)\cite{jovanovic2015subaru}, the Spectro-Polarimetric High-contrast Exoplanet Research instrument (SPHERE)\cite{beuzit2008sphere}, the Gemini Planet Imager (GPI)\cite{macintosh2014first}, the Keck Planet Imager and Characterizer (KPIC)\cite{jovanovic2019keck}, and most recently the Magellan Extreme Adaptive Optics System (MagAO-X)\cite{males2020magao}. All of these instruments have common features: a high actuator count deformable mirror running at extreme speeds (at least 1kHz), a high-performance wavefront sensor (WFS), and a high-contrast coronagraph. These instruments are pathfinders for ExAO systems on the GSMTs.

Detector size, speed, and noise all impose limits on WFS performance. The sampling of the wavefront limits the number of modes for which the adaptive optics (AO) system can correct. The read-out speed of the detector sets an upper limit on the speed of the AO loop. Detector noise degrades the measurement of the wavefront error, forcing slower loop speeds to compensate, resulting in poor correction when using faint guide stars. Higher wavefront sampling requires a larger format detector which takes longer to readout. This trade-off motivates the need for a new wavefront sensor architecture to build first-light ExAO instruments for the GSMTs.

%  The four sided PWFS (4PWFS) is the only type of pyramid used on sky in adaptive optics systems today, but the four sided architecture is not necessary, only three copies of the telescope pupil are needed to compute the X and Y wavefront slopes. The 3PWFS has advantages over the 4PWFS. It is much easier to fabricate a high quality three-sided pyramidal optic, which lowers the cost of the adaptive optic system. A 3PWFS uses less detector pixels than the 4PWFS, making it less sensitive to read noise and possibly a better wavefront sensor than the 4PWFS. 

% A four sided glass pyramid is used to split the focal plane four ways, resulting in four copies of the telescope pupil imaged onto one detector. Intensity fluctuations in the pyramid pupils are related to phase aberrations, and a centroiding algorithm is used to the wavefront slopes. The number of modes controlled in the AO loop is at most equal to the number of pixels that sample a pupil. FIGURE

% INSERT PYRAMID FIGURE AND USE TO EXPLAIN OPERATION

 The GSMTs all plan to use pyramid wavefront sensors (PWFS) in ExAO instruments. The PWFS acts like a Foucault test in 2 dimensions. Light from the telescope is focused onto a glass pyramid tip where it is split and then the pupil plane is reimaged onto a detector. The result is copies of the telescope pupil that contain intensity fluctuations that are related to the wavefront phase. All current PWFS on telescopes use a four-sided pyramid (4PWFS), resulting in four pupil copies. The Planetary Systems Imager \cite{fitzgerald2019planetary} for the TMT will use a non-modulated PWFS\cite{guyon2018wavefront} in combination with lower-order wavefront control to reach and maintain high contrast. The Multi-AO Imaging Camera for Deep Observations (MICADO)\cite{davies2018micado} for the E-ELT is a pathfinder instrument for performing high contrast imaging on GSMTs that uses a PWFS. The High Angular Resolution Monolithic Optical and Near-infrared Integral field spectrograph (HARMONI)\cite{neichel2016adaptive}, also for the E-ELT, has a PWFS in the single-conjugate adaptive optics (SCAO) mode to deliver diffraction-limited performance for the E-ELT's core spectroscopic capability. The Giant Magellan Extreme Adaptive Optics System (GMagAO-X)\cite{males2019gmagao} is being developed as a first light ExAO instrument for the GMT and will use a PWFS.

 As an alternative to the conventional 4PWFS, we are exploring a three-sided PWFS (3PWFS) as a GSMT-ExAO wavefront sensor to reduce the required number of detector pixels. The 3PWFS uses fewer detector pixels than the 4PWFS, and therefore should be less sensitive to read noise. However, the 3PWFS has not been well studied, and no 3PWFS has been tested in a closed-loop adaptive optics system. The goal of this paper is to analyze the refractive three-sided PWFS and demonstrate its operation in simulations. We start by deriving a Slopes calculation for 3PWFS signals and test the validity in a real adaptive optics closed-loop system on the LOOPS\cite{janin2019adaptive} test bench at the Laboratoire d'Astrophysique de Marseille. Then we use the diffraction theory of the Foucault knife edge test to predict the expected signal and sensitivity of the PWFS to Fourier modes. Finally, we use the Object Oriented MATLAB Adaptive Optics toolbox (OOMAO)\cite{OOMAO} to simulate an end-to-end model of an adaptive optics system using a PWFS with modulation and compare the performance of the 3PWFS to the 4PWFS.

\section{The Pyramid Wavefront Sensor}

\begin{figure}
    \centering
    \includegraphics[width=0.75\textwidth]{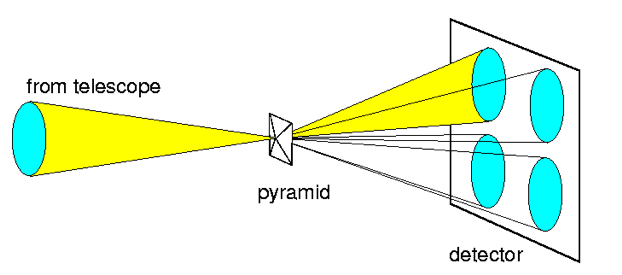}
    \caption{The optical design of a PWFS. Light from the telescope is focused onto a glass pyramid tip where it is split and then the pupil plane is reimaged onto a detector. The result is copies of the telescope pupil that contain intensity fluctuations that are related to the wavefront phase. \cite{pyramidfig}}
    \label{fig:pyramid}
\end{figure}

The optical design of the PWFS consists of four components: a focusing optic, the glass pyramid, and a relay lens to image the pupils on the detector.\cite{ragazzoni2002pyramid} Figure \ref{fig:pyramid} shows a schematic of the operation of a PWFS.  The starlight is first brought to a focus on the tip of a glass pyramid that splits the focal plane into parts. The apex angle of the pyramid imparts a tilt to separate each of the sections. The result is copies of the telescope pupil that are separated spatially on a detector. The number of pixels across each pupil determines the number of modes the instrument is sensitive to. Current on-sky adaptive optics systems that have a PWFS use a four-sided PWFS. The Magellan Adaptive Optics System (MagAO)\cite{close2018status}, the Large Binocular Telescope Interferometer (LBTI)\cite{esposito2011adaptive}, and MagAO-X use an achromatic double four-sided pyramid. The SCExAO instrument uses two crossed roof prisms as its pyramid optic. The PWFS is highly sensitive but suffers from a low dynamic range. A modulator is used to increase the linear range of the PWFS. Modulation is achieved by oscillating a piezo-actuator-driven mirror to drive the PSF on the pyramid tip into a circular pattern with a radius quoted in units of $\lambda/D$, where $\lambda$ is the wavelength, and $D$ is the diameter of the entrance pupil. This increases the effective spot size on the pyramid tip which increases the sensor's linearity at the cost of sensitivity.\cite{guyon2005}  

The intensity measured on the PWFS detector must be processed to recover only the intensity signal due to the wavefront phase error. First, the image is dark subtracted and gain corrected to remove detector artifacts. A threshold mask is then applied to mask out any pixels that do not contain a signal. The wavefront sensor response to a flat wavefront is subtracted off to leave only the intensity pattern in each pupil caused by the wavefront phase error. There are two ways we consider handling this intensity to reconstruct the wavefront. The first is the Raw Intensity method (RI), in which the intensity due to only the phase error is used as-is. The second method is the Slopes Maps (SM) method where the intensity in each pupil is combined to estimate the sine of the wavefront derivative, or slope.\cite{verinaud2004nature} The equation for the 4PWFS is,

\begin{eqnarray}
    S_x=\frac{I_1+I_2-I_3-I_4}{I_1+I_2+I_3+I_4}     \label{4PWFSslopes} \\
    S_y=\frac{I_1-I_2-I_3+I_4}{I_1+I_2+I_3+I_4} \nonumber
\end{eqnarray}

\noindent where $S_x, S_y$ are the local wavefront slopes in the x and y direction, and $I_1...I_4$ are the intensity values of the pixel corresponding to the same location in each pupil. A better way to normalize the $S_x$ and $S_y$ slopes it to normalize by the mean value across all valid pixels on the wavefront sensor detector, instead of normalizing pixel by pixel. This method gives a better slope estimate in low-light conditions and when the beacon brightness is fluctuating. The intensity patterns in the pyramid pupils contain both the X and Y spatial information of wavefront gradient from the diffraction off of the pyramid edges. The RI signal processing method uses the pyramid signal as-is. The benefit of the Raw Intensity method is that it is relatively unaffected by any errors in aligning the pyramid pupils on the detector of a real system. In the Raw Intensity method, each pupil pixel is considered independently in the reconstruction of the wavefront, whereas in the Slopes Maps method a single pixel is referenced to pixels in the same location in the other pupils. Misregistration of the pixels in the Slopes Method results in uncorrected phase errors and so a tighter tolerance on the optical alignment is needed. The major challenge in using the Raw Intensity technique is obtaining a good flat wavefront reference image. The signals in the pyramid pupils are then extracted into a single column vector of intensity values.

\subsection{Performance Comparison}
All current PWFSs on telescopes use a 4PWFS. ExAO systems need a high sampling of the PWFS pupils to optimize performance, and as a result, require larger detectors. We are interested in exploring the 3PWFS as an alternative PWFS for an ELT-ExAO wavefront sensor to overcome these limitations. The 3PWFS only has three copies of the pupil and therefore uses fewer detector pixels than the 4PWFS and should be less sensitive to read noise. This will result in increased SNR for a detection that is read noise dominated. The SNR equation for a CCD camera used for astronomical measurements is\cite{howell2006handbook},

\begin{equation} 
SNR= \frac{R_\star \times t}{\sqrt{(R \times t)+(RN^2 \times n_{pix})+n_{pix}(N_s+N_D)}}
\label{SNR}
\end{equation}

\noindent where $R$ is the photon arrival rate, $t$ is the integration time, $RN$ is the read noise, and $n_{pix}$ is the number of pixels used for wavefront sensing. The terms $N_s$ and $N_D$ refer to the sources of noise from sky background and dark current.

\begin{figure}
    \centering
    \includegraphics[width=0.8\textwidth]{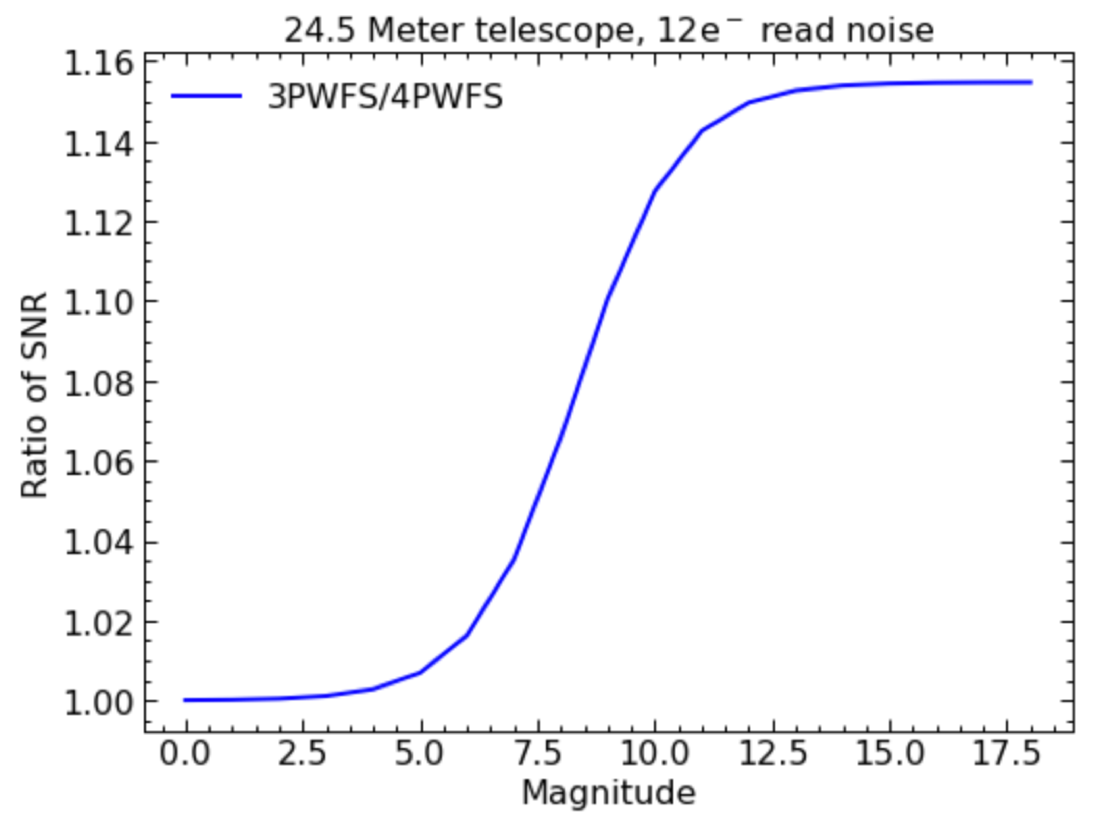}
    \caption{Ratio of the SNR values of the 3PWFS and 4PWFS. In low light conditions when the SNR is dominated by read noise, the ratio converges to $\sqrt{\frac{4}{3}}$.}
    \label{fig:SNRcurve}
\end{figure}

Using equation \ref{SNR} we calculate the SNR for the 3PWFS and 4PWFS based on guide star magnitude. In Figure \ref{fig:SNRcurve} the SNR for the 3PWFS and 4PWFS was calculated for different guide star magnitudes, for a 24.5 meter primary mirror, and 0.001 second integration time. The number of pixels in each pupil was assumed to be 20,000. The 3PWFS used a total of 60,000 pixels, and for the 4PWFS 80,000 pixels were used. Taking the ratio of the SNR curves shows the higher SNR from the 3PWFS in low light conditions. In bright light conditions, the dominant error is photon noise, and the SNR for both the 3PWFS and 4PWFS is the same. As the light level dims and read noise starts to dominate, the SNR of the 3PWFS measurement is higher than the 4PWFS and the ratio converges to a value of $\sqrt{\frac{4}{3}}$.

\begin{figure}
    \centering
    \includegraphics[width=1\textwidth]{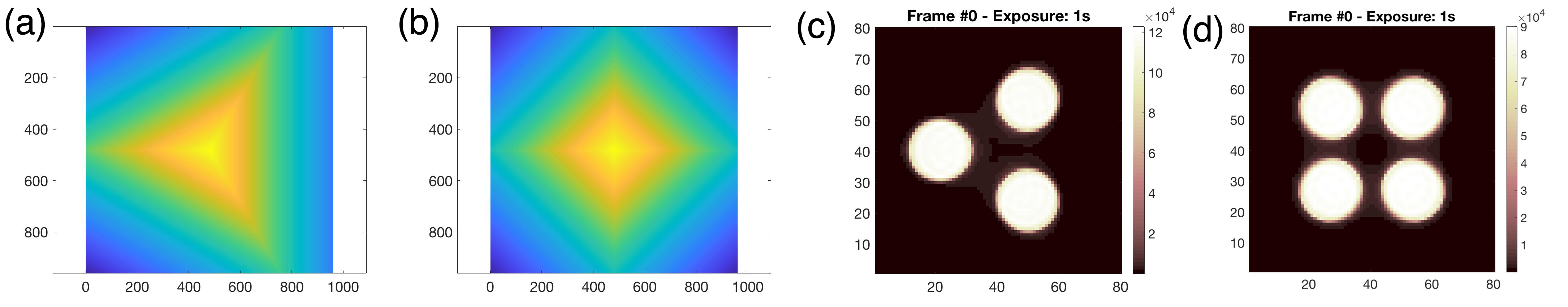}
    \caption{Details of the simulated PWFS in OOMAO. A) and B) The simulated 3PWFS and 4PWFS pyramid masks in OOMAO. These masks are used as a phase screen in the pupil plane to emulate the focal plane splitting and separation done by a real glass pyramid. C) and D) The pupils from the 3PWFS and 4PWFS respectively on the simulated detector. In OOMAO the user can change the size, separation, and intensity values of the pupils through user-defined inputs.}
    \label{fig:oomaoFigs}
\end{figure}

The 3PWFS pupils do not have a natural alignment geometry on a pixel grid detector. The misregistration between pixels in the same location of each of the pupils degrades the performance of the 3PWFS correction when using the Slopes Maps method. The 3PWFS additionally has the potential for a larger null space than the 4PWFS as the wavefront is reconstructed using three points of measurement instead of four. To assess the performance of the 3PWFS compared to the 4PWFS, a simulation was developed using the Object Oriented Matlab Adaptive Optics toolkit (OOMAO) \cite{OOMAO}. The OOMAO toolkit is an end-to-end adaptive optics model that can simulate different combinations of guide stars, turbulent atmospheres, wavefront sensors, deformable mirrors, and science cameras. Light is propagated using Fraunhofer diffraction. The PWFS is simulated using a single tip/tilt phase mask that is segmented into N parts. Figure \ref{fig:oomaoFigs}.A and \ref{fig:oomaoFigs}.C show the masks for the 3PWFS and 4PWFS. The masks are scaled and rotated according to a user input for rotation and pyramid apex angle that controls the separation of the pyramid pupils. After scaling, the mask is converted into a phase mask that is applied at the focal plane to simulate a pyramid tip. Figure \ref{fig:oomaoFigs}.C and \ref{fig:oomaoFigs}.D shows the resulting pyramid pupils on the simulated wavefront sensor camera, using a flat wavefront and 5 $\lambda/D$ modulation. The master OOMAO toolbox simulates a 4PWFS using Slopes Maps. We extended the PWFS class in OOMAO to include a 3PWFS, taking care that the amplitude of the tip/tilt phase in the 3PWFS phase screen generation matched that of 4PWFS. We included our derivation of the Slopes Maps equation for the 3PWFS (derived in Section~\ref{SlopesDerivation} below) and added the Raw Intensity method for both the 4PWFS and the 3PWFS. For both the Slopes Maps and Raw Intensity method the signal was normalized by the mean value across all valid pixels on the wavefront sensor detector, instead of normalizing pixel by pixel.

The threshold mask that determines which pixels are used for wavefront sensing is determined by a calibration step that generates a PWFS with high modulation and a propagated wavefront that contains no phase error. The result is highly uniform pupils on the detector. The detector image is thresholded to mask out any pixels with a low signal. The thresholded image is then converted into a binary mask, where the pixels that contain the pupil signals have a value of 1, and all other pixels have a value of 0. This mask is common to both the SM and RI methods, assuring that in both cases the same pixels are used for wavefront sensing.

\section{3PWFS Slopes Maps}\label{Slopes}

\subsection{Derivation}\label{SlopesDerivation}

The Slopes Maps technique was derived for the 4PWFS using the same intensity centroid calculation as a Shack-Hartmann wavefront sensor. The Shack-Hartmann uses a quad-cell intensity calculation to track the movement of spots to calculate the X and Y gradient of the wavefront phase in post-processing. In the geometric optics approximation, the PWFS acts as a slope sensor, and the Slopes Maps calculates the wavefront gradients in a similar way to the Shack-Hartmann. In the diffractive optics derivation, the wavefront slopes are calculated by the diffraction of the pyramid edges and encoded directly into the intensity pattern. We do not need to perform a Slopes Maps calculation for the pyramid signal because we are directly measuring a function that is related to the gradient phase error, but it is still beneficial to do so. The Slopes Maps is the natural recombination of the pyramid signals that subtracts off the constant Intensity pattern, so a reference image is not required. In a closed-loop system, the wavefront signal is driven towards zero slopes which arises when the PSF is centered on the pyramid tip with no phase errors. The Slopes Maps technique suffers from misalignments of the pyramid pupils; any misregistration in the sampling of the pupils with respect to each other results in a loss of performance. 

The Slopes Maps for the 4PWFS are already known. In previous work by Costa \cite{buchler2004development}, a Slopes Map calculation for the 3PWFS was derived assuming a geometric optics approximation of the PWFS as a derivative wavefront sensor in the modulation regime. In this paper, we introduce a new Slope Maps method to handling the 3PWFS signals, which is derived using the intensity centroid of an equilateral triangle.

\begin{figure}
\begin{center}
\begin{tabular}{c}
\includegraphics[height=5.5cm]{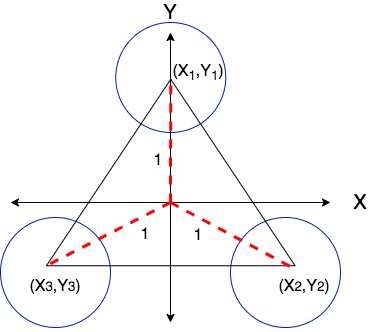}
\end{tabular}
\end{center}
\caption 
{ \label{fig:triCentFig}
Equilateral triangle with the center at the origin. Solving for the coordinates of the vertex points gives the weights for the pixel Intensities used in the Slopes Maps calculation. The blue circles represent the layout of the pupils on the detector. } 
\end{figure}

To calculate the Slopes Maps equations for the 3PWFS, we start from the same assumptions used for the 4PWFS. We assume that the maximum sensitivity of the sensor occurs when the PSF is centered at the tip of the four pyramid edges. We seek a Slopes Maps equation that drives the wavefront to zero slope, resulting in uniform intensity in the re-imaged pupils. To derive the centroid equation for the 3PWFS we start with an equilateral triangle, shown in Figure \ref{fig:triCentFig}. The center of the triangle is at the origin of a Cartesian coordinate system, and the vertices of the triangle are all an equal distance from the origin. The blue circles represent the layout of the pupils on the PWFS detector, and $I_1$ corresponds to $(x_1, y_1)$, etc. Solving for the coordinates $(x_1,y_1),(x_2,y_2), (x_3,y_3)$ gives the weights in the $S_x$ and $S_y$ Slopes Maps. 

% We can see in Equation~\ref{4PWFSslopes} that when the pixel values are equal the value of the Slopes are zero.

For the 3PWFS we find that the $S_x$ and $S_y$ Slopes Maps are,
% \begin{equation}
%     S_x=\frac{\frac{\sqrt{3}}{2}I_2-\frac{\sqrt{3}}{2}I_3}{I_1+I_2+I_3}, \; \;
%     S_y=\frac{I_1-\frac{1}{2}I_2-\frac{1}{2}I_3}{I_1+I_2+I_3}
%     \label{3PWFSslopes}
% \end{equation}

\begin{eqnarray}
    S_x=\frac{\frac{\sqrt{3}}{2}I_2-\frac{\sqrt{3}}{2}I_3}{I_1+I_2+I_3} \label{3PWFSslopes} \\
    S_y=\frac{I_1-\frac{1}{2}I_2-\frac{1}{2}I_3}{I_1+I_2+I_3} \nonumber
\end{eqnarray}

The $S_x$ slopes only depends on the intensities from the $I_2$ and $I_3$ pupil pixels which act as a roof sensor. The result is that care must be taken to properly index the pupils when applying the Slopes Map equation. 
 
 \subsection{Closed-Loop Testbed Verification of the 3PWFS}
 
To test the validity of the Slopes Maps equation for the 3PWFS, and the Raw Intensity method we implemented these signal processing methods into a real closed adaptive optics system using the LOOPS testbed at the Laboratoire d'Astrophysique de Marseille (LAM). The source for the LOOPs is a HeNe laser attenuated by a continuously variable gradient ND filter wheel. There are multiple paths through the loops testbed created by beamsplitters. Different branches can be blocked or passed to create a path where only flat mirrors are illuminated to record a reference PSF or to illuminate only the deformable mirror or phase screen. The phase screen is a milled reflective mirror and simulates a $D/r_0$ of about 28. The wind speed of the turbulence can be changed by speeding up or slowing down a stepper motor that rotates the phase screen. An ALPAO 69 actuator deformable mirror is used for the closed-loop correction. A phase screen applied on a spatial light modulator (SLM) is used to create the pyramid tip. The wavefront sensing camera is an OCAM$^2$K used in single-pixel mode. Each pyramid pupil is 80 pixels in diameter. A full description of the LOOPS testbed can be found in a previous paper by Janin et al.\cite{janin2019adaptive} 

\begin{figure}
    \centering
    \includegraphics[width=0.8\textwidth]{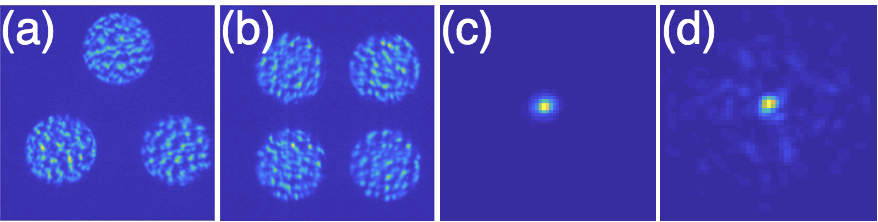}
    \caption{Example images from the LOOPS test-bed of A) the pyramid signal of the 3PWFS under turbulence in closed-loop, B) the pyramid signal of the 4PWFS under turbulence in closed-loop, C) the LOOPS PSF with no turbulence, and D) the LOOPS PSF in closed-loop with the 3PWFS using the Slopes Map technique.}
    \label{fig:LOOOPS}
\end{figure}

On the testbed, a closed-loop correction was achieved for both a 3PWFS and 4PWFS with reconstructors calculated from both the Raw Intensity and Slopes Maps signal handling techniques. The loop was run at 250Hz, with a modulation radius of 5 $\lambda/D$. In all cases, a stable closed-loop was obtained. Figure \ref{fig:LOOOPS} shows example images from the LOOPS test-bed of (A) the pyramid signal of the 3PWFS under turbulence in a closed-loop, (B) the pyramid signal of the 4PWFS under turbulence in a closed-loop, (C) the LOOPS PSF with no turbulence, and (D) the LOOPS PSF in a closed-loop with the 3PWFS using the Slopes Map technique.

Further testing and analysis of the 3PWFS was performed on the Comprehensive Adaptive Optics and Coronagraph Test Instrument (CACTI) at the University of Arizona. Details of the experimental setup and results from a performance comparison test between the 3PWFS and 4PWFS are discussed in an upcoming paper expected late 2021, (Schatz et al., 2021b).

\section{Diffraction Theory of the Foucault Test}\label{diffraction}
\subsection{Derivation of Expected Signal from the Foucault Test}\label{deriv}
  The PWFS is an extension of the knife edge test: each of the pyramid pupils is the signal from a knife edge test, but includes a mirroring effect of the signal across the facets. The underlying physical processes of the pyramid and the knife edge are the same, so we can use the equations that describe the knife edge test to explore the operation of the PWFS. The diffraction theory of the PWFS has been studied in detail. Vérinaud\cite{verinaud2004nature} extends the knife edge diffraction theory to model the signal of a roof sensor with dynamic circular modulation. Hutterer et. al.\cite{hutterer2019real} extends the knife edge analysis into general operators such that modulations of any pattern can be modeled. Shatokhina et. al.\cite{shatokhina2014fast} presents simplified equations to approximate pyramid signals with and without modulation. Correia et. al.\cite{correia2020performance}, and Fauvarque et. al.\cite{fauvarque2019kernel} model the pyramid operation as a Fourier filter. In this paper, we use a formalism similar to Vérinaud and limit our analysis to that of a one-dimensional knife edge test for simplicity. We use the knife edge analysis to derive the relationship of the PWFS sensitivity to Fourier modes. Fourier modes have a direct relationship to locations on the focal plane; which is important for coronagraphy because we are trying to maximize the contrast in the dark hole region created by the coronagraph.
 
 In this section, we consolidate the diffraction theory of the knife edge test by Linfoot\cite{linfoot1948theory}, Katzoff\cite{katzoff1971quantitative}, and Wilson\cite{wilson1975wavefront} into a single derivation with uniform notation. These three papers build upon the same derivation, and hereafter we refer to the collection of papers as LKW. We then deviate from their derivation to link phase aberrations in the shape of Fourier modes to intensity patterns produced by the knife edge test. We assume a focal plane bisected by a binary amplitude mask representing the knife edge. 
 
\begin{figure}
     \centering
     \includegraphics[width=0.9\textwidth]{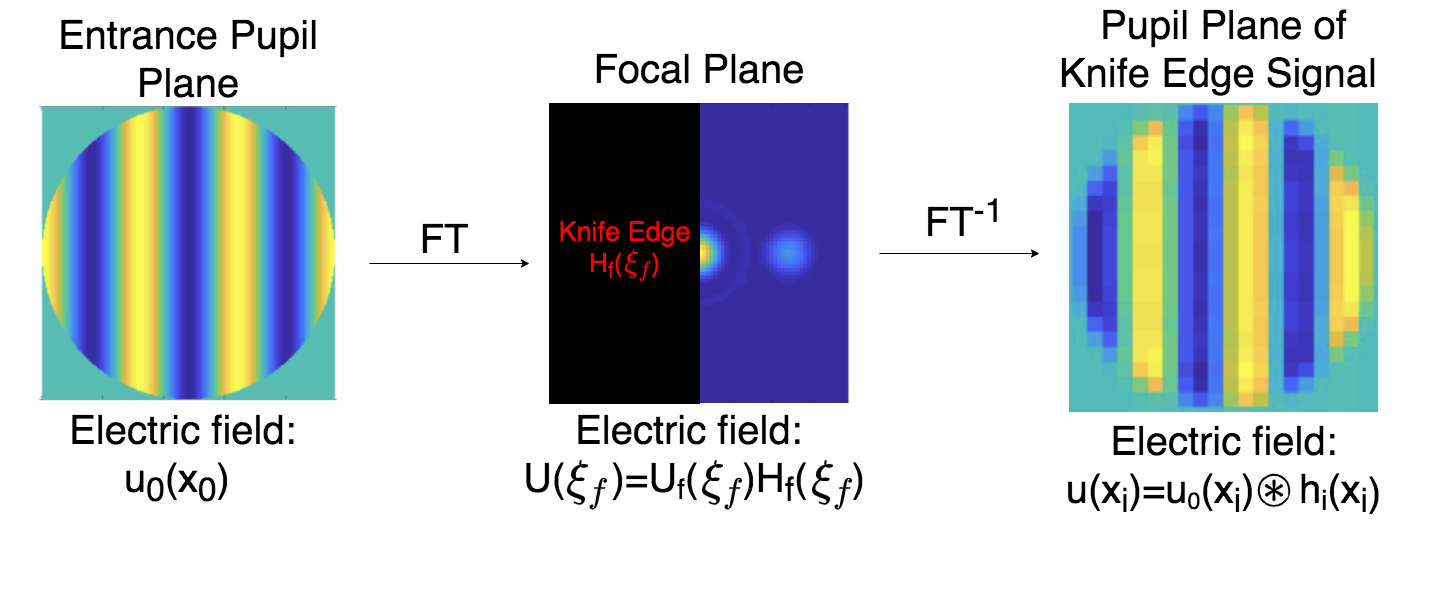}
     \caption{Diagram of the steps taken in the derivation. The entrance pupil with electric field $u_0(x_0)$ is Fourier transformed to the focal plane. At the focal plane the knife edge by multiplying the binary mask function of the knife edge, $H_f(\xi_f)$, by the Fourier transform of the electric field in the entrance pupil. An inverse Fourier transform taken to return to the pupil plane where the pyramid signal is detected.}
     \label{fig:derivationFlow}
 \end{figure}

Figure \ref{fig:derivationFlow} describes the steps taken in this derivation. The diagram is in two dimensions for visualization, but in this derivation, we assume a one dimensional case for simplicity. We start with an electric field in the entrance pupil, $u_0(x_0)$ that has a phase error given by $\cos(nx)$. A Fourier transform is taken to propagate the electric field to the focal plane where the knife edge, $H_f(\xi_f)$ is applied. We don't solve for $U(\xi_f)$, the electric field in this focal plane, and instead, take an inverse Fourier transform to the conjugate pupil plane where the knife edge test signal is found. The electric field at this pupil plane is given by $u(x_i)$, is the convolution of the electric field in the entrance pupil propagated to this pupil plane, $u_0(x_i)$, and the inverse Fourier transform of the knife edge, $h_i(x_i)$. The phase error in the electric field is expanded in a power series, then linearized, and the modulus squared is taken to find the field intensity. By subtracting the constant term, we are left with the intensity pattern due to the phase errors only.

% To start, we define some terms used in the derivation. 

% \begin{center}
%     \begin{tabular}{ | l |p{12cm}|}
%     \hline
%     $u_0(x_0)$ & Complex wavefront at the "phase object". This could be the surface of a mirror under testing, but for our application it is the phase errors in the pupil caused by atmospheric turbulence. \\ \hline
%     $u_i(x_i)$ & Complex wavefront at the image plane. For the pyrmaid wavefront sensor the image plane is the pupil plane. \\ \hline
%     $h(x_i)$ & The impulse response of the system. \\ \hline
%     $m(x)$ & The equation of phase error to be measured. $u(x)=exp(-2\pi i m(x))$  \\ \hline
%     $p(x_i )$ & The intensity pattern at the image plane due to the diffraction of the electric field with the knife edge \\ \hline
%     \end{tabular}
% \end{center}

To start we define the electric field in the entrance pupil as $u_0(x_0)$. We take the Fourier transform of the electric field to get to the focal plane, where we will apply the knife edge. The Fourier transform is here defined as:

\begin{equation}
    \mathcal{F}\{\xi_x\}=\int_{-\infty}^\infty e^{-i 2 \pi x \xi_x} f(x) dx
    \label{FourierT}
\end{equation}

where $\xi_x$ is the spatial frequency corresponding to the spatial coordinate $x$.\cite{gaskill1978linear} $\mathcal{F}\{\cdot\}$ denotes the Fourier transform operator and  $\mathcal{F}^{-1}\{\cdot\}$ will denote its inverse. The equation of the electric field in the focal plane, $U(\xi_f)$ is:

\begin{equation}
    U (\xi_f)=U_f(\xi_f)H_f(\xi_f)
    \label{FT}
\end{equation}

where,
% We now multiply the transmission function of the knife edge, $H(\xi_x)$. In this derivation we assume that the knife edge is like a step function, the transmission of light is 1 until the edge of the knife edge is reached and then drops to zero. We assume for simplicity that the aperture is infinite, but that the knife edge eclipses half of it. After multiplication, the Fourier transform of equation must be taken again to calculate the electric field in the pupil plane. We write out the step function in terms of the sign function $sgn(\xi_x)$, in order to calculate the Fourier Transform easily. The transfer function of the knife edge is given by the following equation.
\begin{equation}
H_f(\xi_f) = 1/2+1/2\mbox{sgn} \begin{cases} 1, & \mbox{for  } \xi_f>0 \\ 0, & \mbox{for  } \xi_f<0 \end{cases}
\end{equation}
% \[ 
% H_f(\xi_f)= 1/2+1/2\mbox{sgn}(\xi_f) \left\{
% \begin{array}{ll}
%       1 \: for \: \xi_f >0\\
%       0 \: for \: \xi_f <0\\
% \end{array} 
% \right 
% \]   

is the binary mask function of the knife edge, and $U_f(\xi_f)$ is the Fourier transform of $u_0(x_0)$. Taking the inverse Fourier transform brings us to the conjugate pupil plane. The inverse Fourier transform of $H_f(\xi_f)$ is

% \begin{equation}
%      FT^{-1}[H(\xi_x)]=\frac{1}{2\pi}\int_{-\infty}^\infty e^{ i x \xi_x} ( \frac{1}{2}+\frac{1}{2}sgn(\xi_x)) d\xi_x \\
%  =\frac{1}{2}\delta(-x)+\frac{i}{2\pi}(\frac{1}{x})
% \label{delta}
% \end{equation}

\begin{equation}
    h_i(x_i)= \mathcal{F}^{-1}[H_f(\xi_f)] =
    % \frac{1}{2\pi} \int_{-\infty}^\infty e^{ i x \xi_f}
    % \left(
    %     \frac{1}{2} + \frac{1}{2} \mathrm{sgn}(\xi_f)
    % \right) d\xi_f
    % = 
    \frac{1}{2} \delta(-x_i) + \frac{i}{2\pi} \left(\frac{1}{x_i}\right)
\label{delta}
\end{equation}

%  = \frac{1}{2} \frac{1}{2\pi} \int_{-\infty}^\infty e^{- i x \xi_x}d\xi_x+\frac{1}{2} \frac{1}{2\pi}\int_{-\infty}^\infty e^{- i x \xi_x}sgn(\xi_x) d\xi_x
% \end{align*}

% We make use of the property of the delta function to solve for the Fourier Transform of $\frac{1}{2}$:

% \begin{equation}
%     \delta(x)=\frac{1}{2\pi}\int_{-\infty}^\infty e^{ i x \xi_x}d\xi_x
% \end{equation}

% which leads to,
    
% \begin{equation}
%     \frac{1}{2}\delta(x)=\frac{1}{2}\frac{1}{2\pi}\int_{-\infty}^\infty e^{ i x \xi_x}d\xi_x 
% \end{equation}

% where $\frac{1}{2}\delta(x)$ is the Fourier Transform of $\frac{1}{2}$. To solve for the Fourier transform of $\frac{1}{2} sgn(\xi_x)$ function, we use the derivative property.

% \begin{equation}
%   FT[\frac{d g(x)}{dx}]=i2\pi \xi_x G(\xi_x)
%   \label{derivativeProp}
% \end{equation}

% where the derivative of $\frac{1}{2} sgn(\xi_x)$ is $\delta(\xi_x)$. Plugging this into Equation~\ref{derivativeProp}, gives the result,
% \begin{equation}
%   FT[\frac{d sgn(\xi_x)}{d\xi_x}]=FT[\delta{\xi_x}]=1=2\pi i x G(x).
% \end{equation}

% Solving for $G(x)$,
% \begin{equation}
%  G(x)=\frac{1}{i2\pi x}.
% \end{equation}

% Now we can sum our answers to get the Fourier Transform of a step function.

% \begin{equation}
%  FT[step(\xi_x)]=\frac{1}{i2\pi x_i}+\frac{1}{2}\delta(x_i)
%  \label{knifeedge}
% \end{equation}

We take the results of Equation~\ref{delta} and convolve it with $u_0(x_i)$, which is the Fourier transform of $U_f(\xi_f)$ in Equation~\ref{FT}, to get the electric field at the pupil plane, $u(x_i)$. The resulting equation is Equation 5a from Wilson.\cite{wilson1975wavefront}

% \begin{equation}
%   u_i(x_i)= u_0(x_0) \circledast \frac{i}{2\pi x_i}+\frac{1}{2}\delta(-x_i) 
%  = \int_{-\infty}^\infty u_0(x')[\frac{1}{2}\delta(x'-x_i)+\frac{i}{2\pi}\frac{1}{x_i-x'}]dx' 
% \end{equation}
%  \begin{equation}
%       u_i(x_i) =\frac{1}{2}u_0(x_i)+\frac{i}{2\pi}\int_{-\infty}^\infty \frac{u_0(x')dx'}{x_i-x'}
%       \label{KEconv}
%  \end{equation}

\begin{equation}
    u(x_i)= u_0(x_i) \circledast \left(
        \frac{i}{2\pi x_i}+\frac{1}{2}\delta(-x_i)
    \right)
    =
    \int_{-\infty}^\infty
    u_0(x') \left[
        \frac{1}{2}\delta(x'-x_i)+\frac{i}{2\pi}\frac{1}{x_i-x'}
    \right] dx' 
    \label{KEconv}
\end{equation}

In high contrast imaging, we are interested in the effect of phase errors at different spatial frequencies as these correspond to the contrast at specific locations in the post-coronagraphic focal plane. To find the response of a knife edge to an error of the form $\cos(nx)$, where $n$ is the spatial frequency, we need to modify the derivation by LKW to reach a form that allows us to easily compute the expected signal of the Foucault test from a phase error in the form of a Fourier mode. We make two modifications to LKW: we assume the infinity pupil approximation, and consider only the linear component of the electric field. The assumptions made by LKW lead to an integral that describes the intensity signal of a knife edge test that is difficult to solve. Katzoff attempts a modified approach for solving the integral for phase errors in different forms. This approach did not produce a good match for the intensity signal from a phase error in the shape of a Fourier mode. This leads us to revisit the derivation and modifying the assumptions to better reflect the knife edge approximation to the pyramid wavefront sensor. In the derivation by LKW the infinite limits of the integral are set arbitrarily to -1 to 1 since the factor $dx'/(x_i-x')$ is nondimensional. \cite{katzoff1971quantitative} They found that the intensity pattern given by the infinite approximation is valid for within the pupil boundaries but gave an infinite irradiance at locations on the pupil edge. We are only interested in the intensity pattern inside the pupil boundary. We assume an infinite aperture approximation and take the integral from $-\infty$ to $\infty$. LKW expand the electric field and consider the quadratic component as well as the linear component of the electric field to the knife edge signal.  We are interested in only the linear component of the knife edge signal. In our expansion of the electric field, we only consider the linear component and substitute the result into Equation \ref{KEconv}.

The electric field at the entrance aperture is now:

\begin{equation}
    u_0(x_i )=\exp(-2\pi i m(x_i ))=1-2\pi i m(x_i).
    \label{EF}
\end{equation}

We substitute Equation~\ref{EF} into Equation~\ref{KEconv} to find the electric field in the pupil plane $u(x_i)$, which yields:

\begin{equation}
    u(x_i)= \frac{1}{2}(1-2\pi i m(x_i))+\frac{i}{2\pi}\int_{-\infty}^\infty \frac{(1-2\pi i m(x'))dx'}{x_i-x'}
    \label{HT}
\end{equation}

Expanding the integral, we have:
\begin{equation}
    u(x_i)= \frac{1}{2}(1-2\pi i m(x_i))+\frac{i}{2\pi}\int_{-\infty}^\infty \frac{dx'}{x_i-x'}+\int_{-\infty}^\infty \frac{ m(x')dx'}{x_i-x'}
\end{equation}

where the integrals are now in the form of a Hilbert transform.\cite{villa2014foucault} The Hilbert transform is defined as:

\begin{equation}
    H(y(t))=\frac{-1}{\pi} PV\int_{\infty}^{\infty} \frac{y(t') dt'}{t'-t}
\end{equation}

which has the property that the Hilbert transform of a constant is 0.\cite{poularikas2018handbook} In this equation $PV$ stands for the principal value. The $PV$ is used to evaluate integrals that have a discontinuity, in our case this is when $x_i=x'$.\cite{johansson1999hilbert} When the wavefront error is given by $m(x)=\cos(nx)$, Equation~\ref{HT} becomes,

\begin{equation}
    u(x_i)= \frac{1}{2}(1-2\pi i \cos(nx_i))+\int_{-\infty}^\infty \frac{ \cos(nx')dx'}{x_i-x'}
\end{equation}

To get the integral in the form of a Hilbert transform we multiply by $\frac{\pi}{\pi}$. The Hilbert transform of $H(\cos(nx))=\cos(nx+\frac{\pi}{2})=-\sin(nx)$.

\begin{equation}
\int_{-\infty}^\infty \frac{ \cos(nx')dx'}{x_i-x'}=\pi*[\frac{1}{\pi}\int_{-\infty}^\infty \frac{ \cos(nx')dx'}{x_i-x'}]=-\pi \sin(nx_i)
\end{equation}

The equation for the electric field in the pupil plane of the knife edge signal is,

\begin{equation}
   u(x_i)= \frac{1}{2}-i \pi \cos(nx_i) -\pi \sin(nx_i)
   \label{derivationResult}
\end{equation}

and we can now take the modulus squared to get the intensity.

\begin{equation}
    I(x_i)=\frac{1}{4}+\pi^2-\pi \sin(nx_i)
    \label{equationIntensityResult}
\end{equation}

Subtracting the constant terms which represent the intensity response $I_r (x_i )$ to a perfect wavefront, gives the intensity pattern due to the phase errors only. In this derivation, we find an intensity pattern that is proportional to $-\sin(nx)$ resulting from a phase error of the form $\cos(nx)$. In the next section, we demonstrate that this is the expected response.

The diffraction theory of the Foucault test predicts an intensity pattern that is the Hilbert transform of the wavefront phase. The result is an intensity pattern that is related to the wavefront derivative but is not quite the derivative. In the case of a derivative wavefront sensor such as the Shack-Hartmann, the operation of the wavefront sensor on the phase to signal measurement is $\frac{d}{dx}\cos(nx)=-n\sin(nx)$. In the case of the pyramid, the reference subtracted result is  $\frac{d}{dx}\cos(nx)=-\sin(nx)$, which is similar to the derivative wavefront sensor but without the scaling by spatial frequency. The result is that the knife edge and therefore the unmodulated pyramid sensitivity does not depend on spatial frequency, because all spatial frequencies will produce an equally strong signal. In the Shack-Hartmann case, the strength of the signal scales linearly with spatial frequency.

\subsection{Verification}

\begin{figure}
    \centering
    \includegraphics[width=0.8\textwidth]{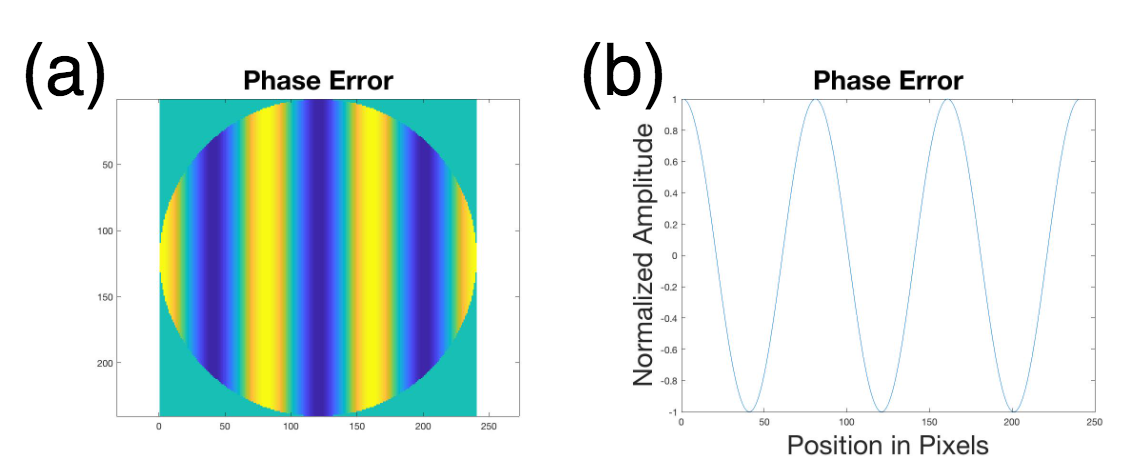}
    \caption{A. Applied cosine phase error in radians in pupil plane to be propagated through the PWFS. B. Scaled cross section of the phase error.}
    \label{fig:CosinePhaseDiagram}
\end{figure}

\begin{figure}
    \centering
    \includegraphics[width=1\textwidth]{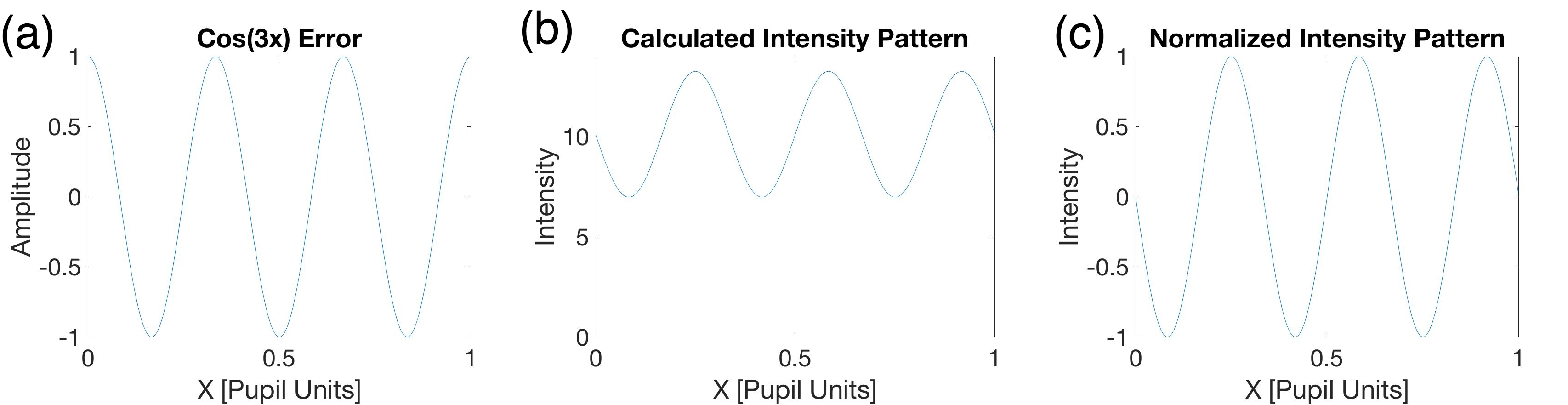}
    \caption{Predictions from the Foucault derivation. A. The applied $\cos(3x)$ phase error. B. The intensity pattern predicted for a single pupil. C. The expected pyramid signal after subtracting constant terms and scaling.}
    \label{fig:MathPredicions}
\end{figure}

The approximations we used to derive Equation~\ref{derivationResult} are supported by our simulation of the response of a PWFS to a cosine phase error. Using OOMAO we apply a Fourier mode as a phase error in the pupil plane, propagate through the pyramid onto the detector, and examine the resulting intensity pattern on the detector as well as the calculated slopes. The derivation in Section \ref{deriv} is an approximation valid when the wavefront sensor is operating in the linear regime.  We first consider the modulated case to ensure the linearity of the pyramid signal. We then consider the unmodulated pyramid to confirm that the derivation is still a good approximation of the unmodulated PWFS signal. The PWFS are under $5 \frac{\lambda}{D}$ modulation to ensure that the pyramid response is linear for the amplitude of Fourier mode applied, and we are not including any noise. We apply one radian of phase error that is in the form of $\cos(3x)$ to both the modulated and unmodulated PWFS. The phase error in radians is shown in Figure \ref{fig:CosinePhaseDiagram}, where Figure \ref{fig:CosinePhaseDiagram}.A is the $\cos(3x)$ phase error, and Figure \ref{fig:CosinePhaseDiagram}.B is a cross-section of the center function in the $x$-direction. We compare the resulting intensity patterns to our prediction in Equation~\ref{equationIntensityResult}. For an intensity pattern of the form $\cos(3x)$ shown in Figure \ref{fig:MathPredicions}.A, we would expect an intensity pattern in a pyramid pupil to be $1/4+\pi^2-\pi \sin(3x)$ displayed in Figure \ref{fig:MathPredicions}.B. After subtracting constant terms we scale the amplitude of the signal such that the maximum value is 1, to compare the resulting pyramid signal is $-sin(3x)$ plotted in Figure \ref{fig:MathPredicions}.C.

\begin{figure}
    \centering
    \includegraphics[width=0.8\textwidth]{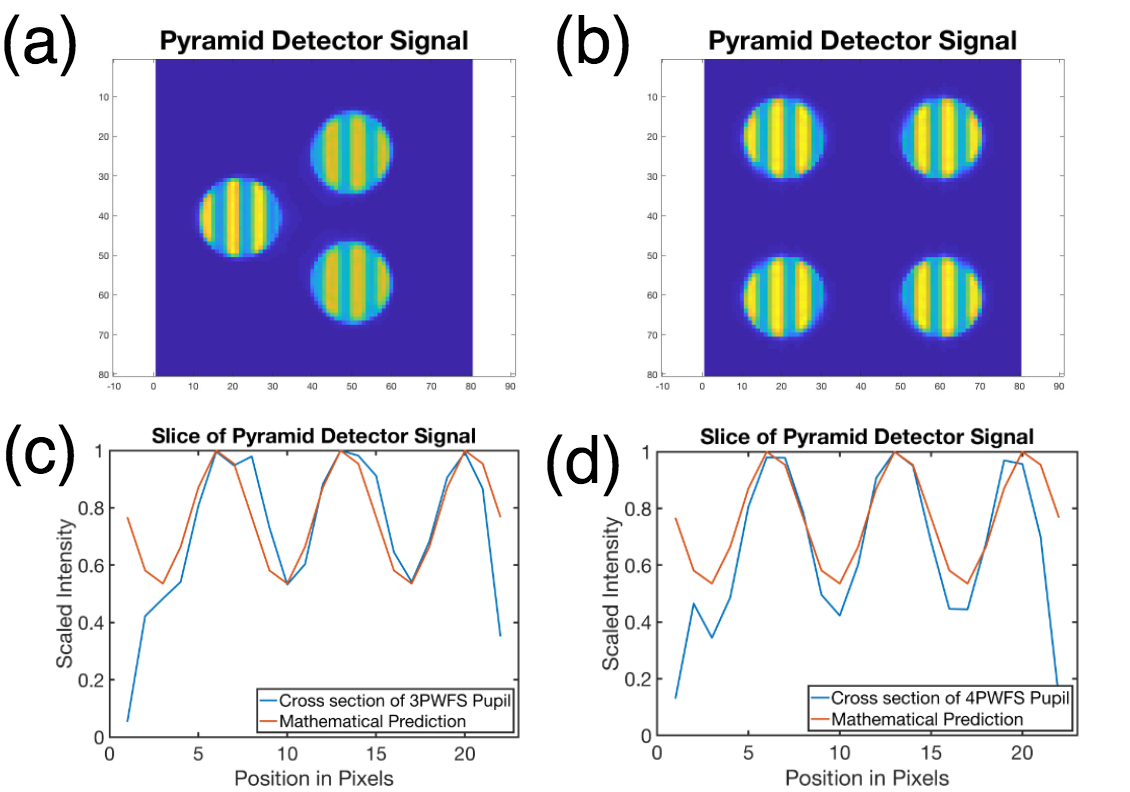}
    \caption{Resulting intensity patterns on the PWFS detector from A. the 3PWFS, and B. the 4PWFS. Figures C. and D. are cross sections of the intensity pattern plotted against the mathematical prediction. For C. The cross-section was taken from center of the middle-left pupil. For D. The cross section was taken from the center of the top two pupils.}
    \label{fig:IntensityPatternsDiagram}
\end{figure}

The resulting intensity patterns from the 3PWFS and 4PWFS are shown in Figure \ref{fig:IntensityPatternsDiagram} A and B. In Figure \ref{fig:IntensityPatternsDiagram} C and D we take a cross-section from one of the pupils from each wavefront sensor and scale the signal. We then plot this against the predicted scaled intensity pattern. We find that the mathematical prediction is a good match for the measured intensity pattern inside the pupil of a modulated pyramid when the signal is linear.

\begin{figure}
    \centering
    \includegraphics[width=0.9\textwidth]{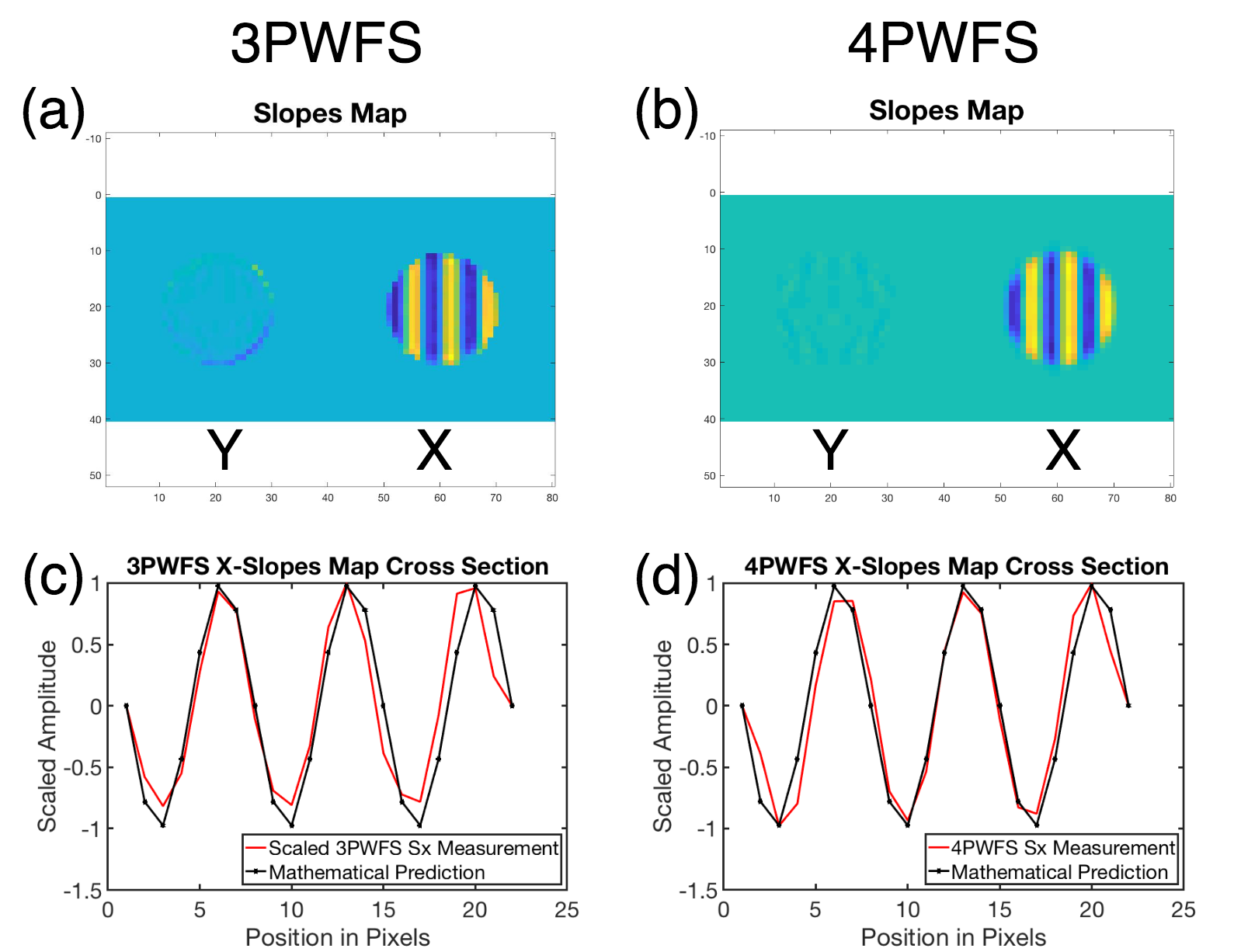}
    \caption{A. and B. Calculated slopes from the pyramid pupils. C. and D. Cross section of the Slopes plotted against the predicted signal. }
    \label{fig:SlopesMapDiagram}
\end{figure}

To gain further clarity we use the Slopes Maps equation that subtracts off the constant intensity and turns the signal into a pure X and Y measurement of phase. The Slopes Maps naturally subtracts the flat wavefront response and in that way is self-referencing.  More details on the Slopes calculation for the 3PWFS are in Section~\ref{Slopes}. The Slopes Maps for the 3PWFS and the 4PWFS are shown in Figure \ref{fig:SlopesMapDiagram}, as well as the cross-sections that are indeed a $-\sin(nx)$ function. 

\begin{figure}
    \centering
    \includegraphics[width=0.9\textwidth]{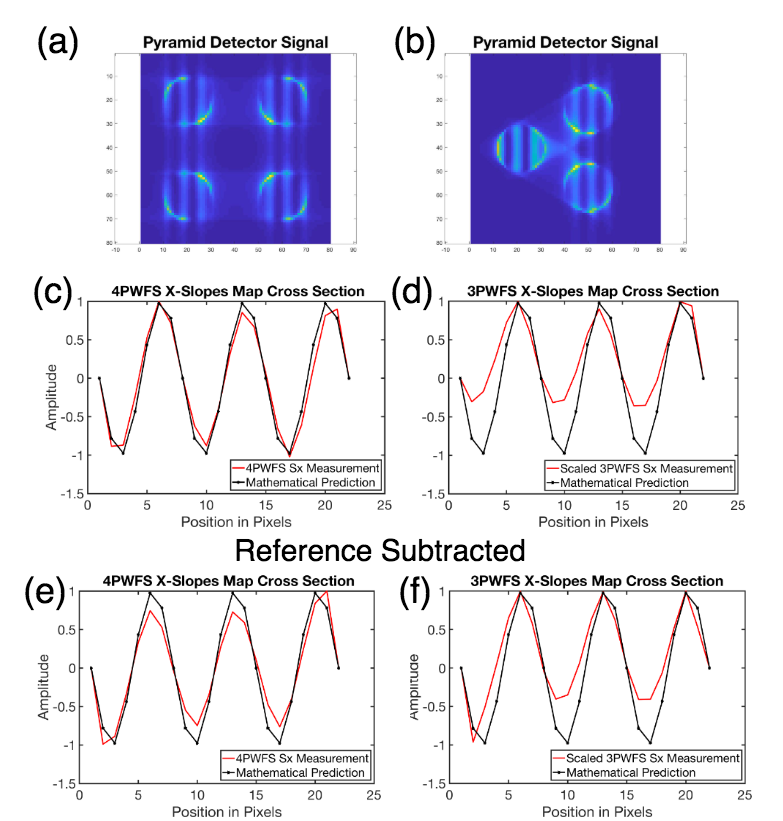}
    \caption{A. and B. The wavefront sensor detector signals. C. and D. cross-section of the wavefront sensor Slopes Maps plotted against the prediction. E. and F. the cross-section of the Slopes Maps with the reference signal subtracted.}
    \label{fig:Mod0}
\end{figure}

The previous simulations were performed with $5\lambda/D$ modulation. We can consider the 0 modulation case, which is a more direct relation to the results from the derivation. The unmodulated PWFS signal is more nonlinear than the modulated pyramid and therefore the fit between the prediction and the measured signal will be poor. In the case of the Slopes Maps signal, we found in simulation that the self-referencing is degraded from a nonlinear signal, and to return to a closer estimate of the signal the flat wavefront reference signal should be subtracted off. The fit worsens for the Slopes Maps without reference subtraction when there is a misregistration of the pupils, as will always be the case for the 3PWFS. For the 3PWFS subtracting a reference improved the mean square error of the fit of simulation to a derivation by about a factor of two, from 0.147 without subtraction, to 0.096 in the reference subtracted case. Figure \ref{fig:Mod0} summarizes the findings from the simulations with 0 modulation. Figure \ref{fig:Mod0}.A and \ref{fig:Mod0}.B are the detector signal of the 4PWFS and 3PWFS to the $\cos(3x)$ phase error. Figure \ref{fig:Mod0}.C and \ref{fig:Mod0}.D are a cross-section of the scaled Slopes Maps plotted against the predicted signal. Figure \ref{fig:Mod0}.E and Figure \ref{fig:Mod0}. F are cross-sections of the Slopes Maps with the reference signal subtracted.

%% INCLUDE A PLOT OF THE -SIN(NX) Pattern as well. Maybe constant +sin for figure 3.

% \begin{figure}[h]
%     \centering
%     \includegraphics[width=0.8\textwidth]{ResultsPlaceholder.png}
%     \caption{ITS JUST A PLACEHOLDER STOP JUDGING ME}
%     \label{fig:sim result}
% \end{figure}

% At the focal plane the diffraction pattern from circular pupil illuminated with a flat wavefront is a sombrero function, which is symmetric. The resulting intensity pattern in the absence of noise, at the detector of a 4PWFS under these conditions, are four pupils each with the same intensity. The sum of the intensities in each pupil is the same, meaning $I_1=I_2=I_3=I_4$.

\section{Performance Comparison of the 3PWFS and 4PWFS}
\subsection{Simulation}

\begin{figure}
    \centering
    \includegraphics[width=0.8\textwidth]{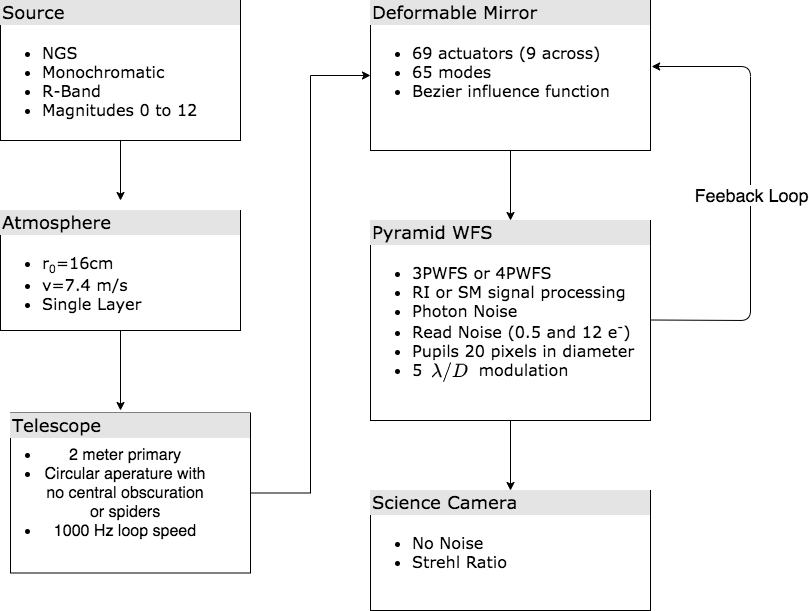}
    \caption{Experimental details of the simulation done in OOMAO. Light starts at the natural guide star and is propagated through the atmosphere, telescope and wavefront sensor. The wavefront sensor measures a correction that is then applied by the deformable mirror.  The resulting AO corrected PSF is recorded on a noiseless science camera. }
    \label{fig:simulationl}
\end{figure}

In Section \ref{diffraction} we derived the response of the Foucault knife edge to a Fourier mode. We then confirmed in simulation that both the 3PWFS and 4PWFS produce a signal that is well described by the predictions of our derivation. These signals are used to reconstruct the wavefront in an AO closed-loop correction. Using OOMAO an end-to-end simulation of an adaptive optics system was written to compare the performance of the 3PWFS and 4PWFS using both Raw Intensity and Slopes.  The goal was to measure the quality of correction from the AO closed-loop by calculating the Strehl Ratio produced by each wavefront sensor as a function of guide star magnitude. The AO correction was optimized by measuring Strehl for each guide star magnitude, and scanning over loop gains from 0.1 to 1.8. In our simulations, we do not include system latency which resulted in an optimum loop gain of over 1 for bright guide stars. A maximum gain of 1.8 was chosen to ensure that the loop gain corresponding to the best correction was chosen. We plotted Strehl versus loop gain for each guide star magnitude as a diagnostic tool to verify that Strehl was decreasing at high loop gains and that we did not need to scan to higher values. In this simulation the Strehl ratio was calculated using OOMAO's built-in Strehl calculator, which uses the OTF calculated from a PSF with no phase aberration, and a PSF with AO compensated phase aberration. The details of the simulation are summarized in the flow chart in Figure \ref{fig:simulationl}. To characterize the sensitivity of each sensor to read noise the following experiment was performed twice, once at 0.5 $e^-$ read noise to match that of an OCAM2K camera, and once at 12 $e^-$ read noise to match the noisiest camera on the Comprehensive Adaptive Optics and Coronagraph Test Instrument (CACTI) at the University of Arizona Extreme Wavefront Control Lab (XWCL). Listed below is the experiment performed in OOMAO:

\begin{itemize}
    \item Guide star magnitude was varied incrementally from 0 to \nth{12} magnitude in steps of 2.
    \item At each magnitude closed-loop data was taken with different loop gains from 0.1 to 1.8 in steps of 0.1. Loop gains of above 1 were used because the simulated system has no latency allowing for higher gains to be used to increase performance.
    \item At each loop gain the performance was tested by closing the loop on 11 different atmospheric realizations and recording 500 closed-loop PSF frames. 
    \item Strehl values were calculated from the 5,500 PSFs and an average Strehl ratio is reported for a given loop gain at a given magnitude.
    \item The loop gain that gave the highest Strehl value for that guide star magnitude was used in the final calculation of Strehl vs guide star magnitude.
    \item The result is a plot of Strehl versus guide star magnitude, where the value of Strehl has been optimized over the loop gain shown in Figure \ref{fig:overall}
\end{itemize}

\subsection{Results}

\begin{figure}[h]
    \centering
    \includegraphics[width=0.8\textwidth]{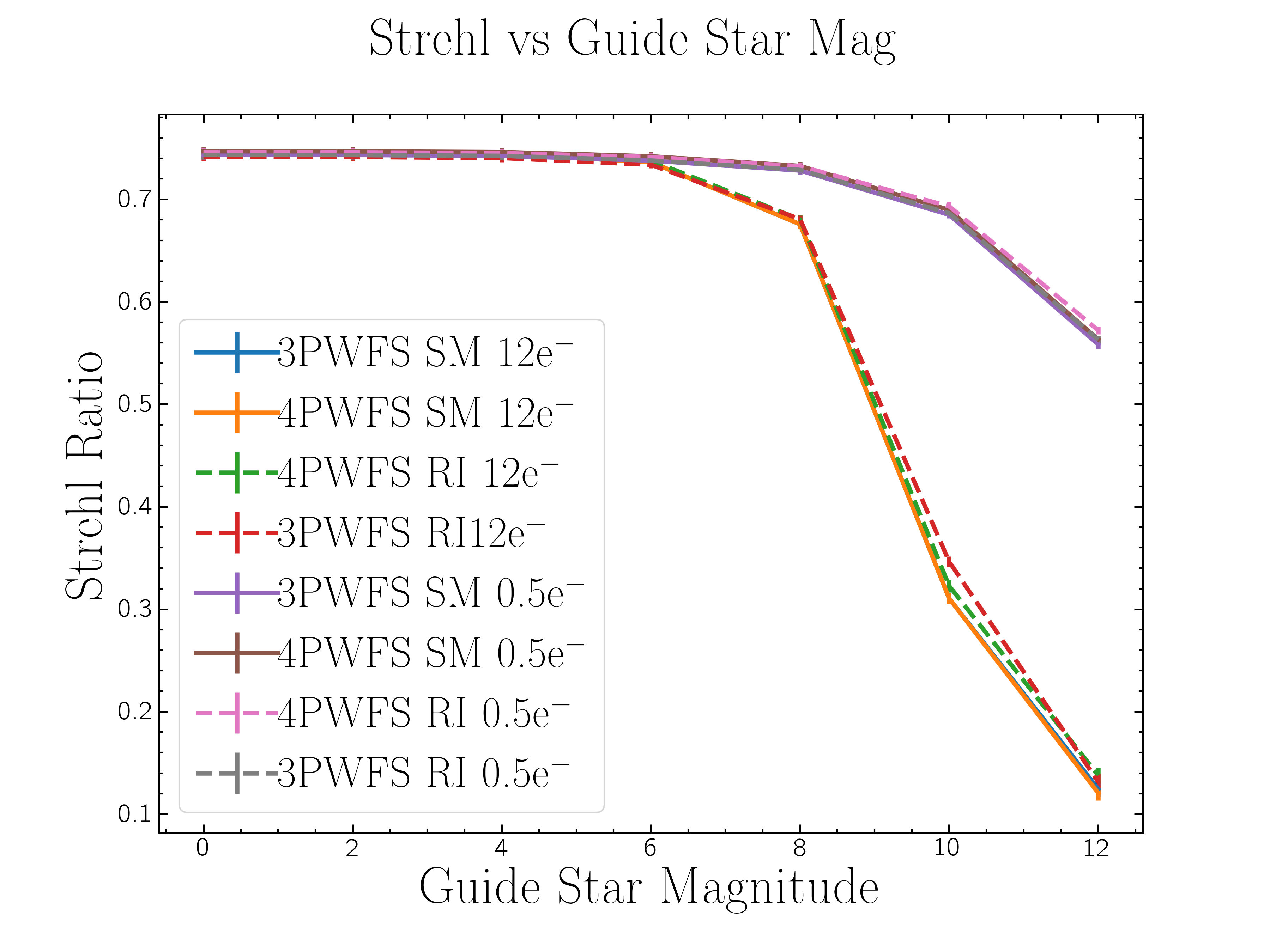}
    \caption{Strehl vs Guide Star magnitude in R-band for each PWFS. The steep drop-off in performance of the lower curves is due to the effects of 12 $e^-$ read noise. At a guide star magnitude of 10, the read noise starts to matter, and the 3PWFS out performs the 4PWFS.}
    \label{fig:overall}
\end{figure}

\begin{figure}[h]
    \centering
    \includegraphics[width=.7\linewidth]{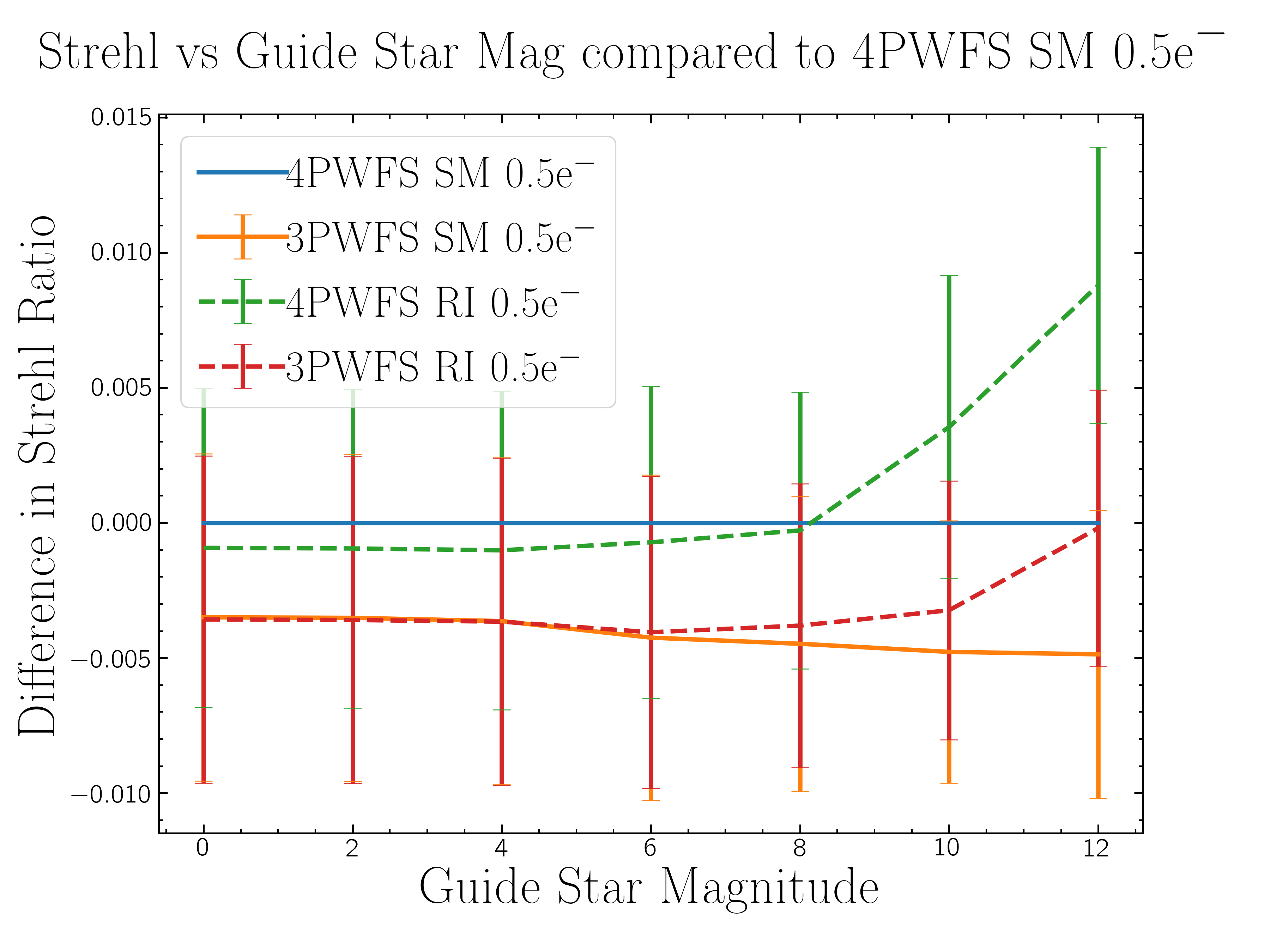}
    \caption{Comparison plot of wavefront sensor performance vs the 4PWFS using Slopes at 0.5 $e^-$ read noise in R-band. }
    \label{fig:0RN}
\end{figure}

\begin{figure}[h]
    \centering
    \includegraphics[width=.7\linewidth]{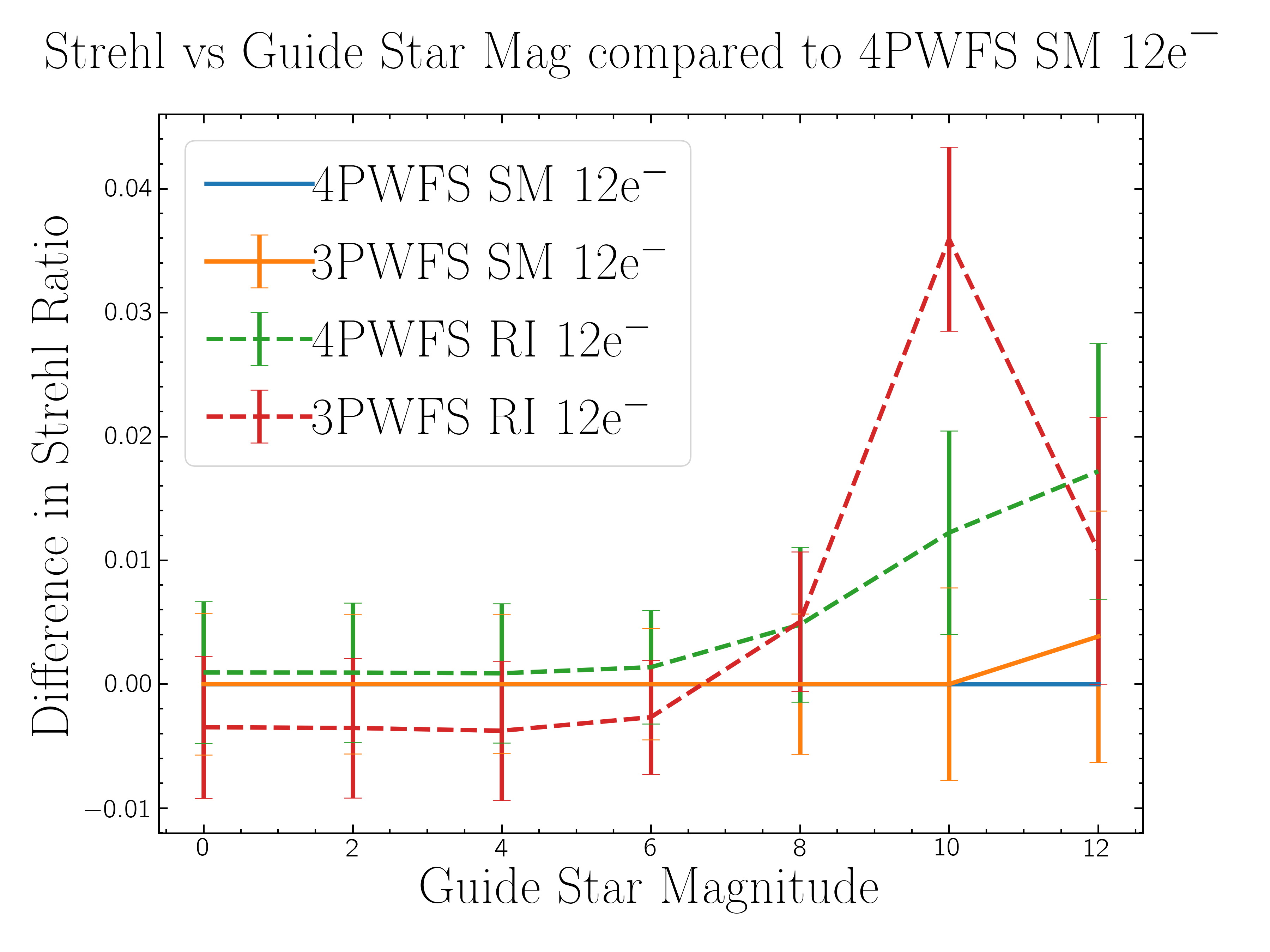}
    \caption{Comparison plot of wavefront sensor performance vs the 4PWFS using Slopes at 12 $e^-$ read noise in R-band.}
    \label{fig:12RN}
\end{figure}

In Figure \ref{fig:overall}, no significant difference in performance was found in the comparison case of the wavefront sensors with 0.5 $e^-$ read noise. Most on-sky adaptive optics systems use a 4PWFS with Slopes (4PWFS SM) so we use this wavefront sensor as our reference for comparison. The performance of each wavefront sensor was found to be within a percent of Strehl from the 4PWFS SM, across all stellar magnitudes. For the simulations at 12 $e^-$ read noise in Figure \ref{fig:overall} an increase of 0.0359 Strehl was found for the 3PWFS using Raw Intensity (3PWFS RI) over the 4PWFS SM at a stellar magnitude of 10. At the same magnitude, the 4PWFS RI also outperformed the 4PWFS SM, but the increase was only 0.0122 Strehl. This simulation successfully showed the increase in performance from the 3PWFS at low light levels where the effects of read noise are stronger. The overall performance of each wavefront sensor at each guide star magnitude is given in Figure \ref{fig:overall}. In Figures \ref{fig:0RN} and \ref{fig:12RN} comparison plots were made by subtracting the Strehl values for 4PWFS SM from the other wavefront sensors, to help visualize the performance of each wavefront sensor. From these results, we can conclude that the 3PWFS is a viable wavefront sensor with performance comparable to a 4PWFS.

\newpage

\section{Discussion}

The diffraction theory of the Foucault test predicts an intensity pattern that is the Hilbert transform. The resulting signal is related to the gradient of the wavefront phase, but without the scaling by spatial frequency. The PWFS is an extension of the Foucault test, and the diffraction theory accurately predicts the intensity patterns seen by a PWFS. Similar to the Foucault test, the sensitivity of a PWFS with no modulation is independent of spatial frequency as found in previous work by Guyon\cite{guyon2005} because all spatial frequencies produce an equally strong signal.

 Codona et al.\cite{codona2018comparative} used the AOsim3 wave optics package to compare the end-to-end performance of a 3PWFS to a 4PWFS with no modulation at 633 nm wavelength. In this simulation, the Raw Intensity signal handling method was used. In open-loop with no noise, the study showed that the 4PWFS outperformed the 3PWFS by 0.005 Strehl. Our simulation results agree with the findings by Codona et al. We found in our simulations with 0.5 read noise that the performance of the wavefront sensors is within 0.01 Strehl. Our results differ from Codona et al. when including read noise. They found that for a detector with 3 $e^-$ read noise, that there was a gap performance improvement by 1 guide star magnitude in favor of the 3PWFS. Codona et. al. considered other factors including system latency. Our simulations are in R band and were optimized only for loop gain. Using a detector with 12 $e^-$ read noise we found that the performance of the wavefront sensors are once again comparable, and we found only $~0.02$ increase in Strehl from the 3PWFS at \nth{10} magnitude. The differences between the two simulations could explain the discrepancies in the results. Codona et al. simulate a smaller telescope aperture that is 1.5m in diameter with a central obscuration. In addition, they have worse seeing conditions with $r_0$=5 cm, but have a larger 12x12 actuator deformable mirror and pyramid pupils that are 34 pixels in diameter. Further work would explore the optimum PWFS design based on the system parameters such as telescope size and seeing conditions. The scope of our simulations is limited but still agree that the 3PWFS is less sensitive to read noise, however, the amount of improvement gained is still under question. Both the 3PWFS and 4PWFS discussed in this paper are refractive, and image all pupils onto the same detector. Moving towards a reflective PWFS that would image each pupil onto its own detector could have a more substantial benefit. Each detector would be smaller, resulting in faster readout speeds with less added read noise. Low-read noise detectors are expensive, so using a reflective PWFS would increase system cost and complexity.
 
Comparing the results from our simulations we see that for 0.5 $e^-$ read noise and 12 $e^-$ read noise results, the RI method has slightly better performance than the SM method for both the 3PWFS and 4PWFS in low light conditions. A possible explanation comes from Section 4.2. A better fit for the Slopes Maps method was found for the unmodulated 3PWFS when a flat wavefront reference was subtracted. A main benefit of the SM method is that it is self-referencing and does not inherently need the flat wavefront reference as these references can be difficult to take. If the PWFS signal does not match the linear response signal expected for a given phase error, due to noise, nonlinearity, or pupil misregistration the self-referencing of the SM method might not be as accurate. This is supported by our results in Figure \ref{fig:0RN} and Figure \ref{fig:12RN} which suggest that the SM method suffers more from the addition to noise as the more significant deviations in performance between the RI and SM method occur at faint guide stars.

 We have found that the 3PWFS is a viable wavefront sensor that is able to fully reconstruct a wavefront and produce a stable closed-loop. Our simulations assumed a perfect wavefront sensor with the absence of manufacturing errors. A high-quality four-sided pyramid optic is difficult and expensive to manufacture. Common errors on four-sided pyramid manufacturing include roofing of the pyramid tip or chipping of the pyramid tip which results in a loss of Strehl.  The 3PWFS is an exciting opportunity for systems that face a trade-off between cost and quality. A real 3PWFS could be cheaper and offer better performance than a 4PWFS. As a next step to developing the potential of the 3PWFS, the University of Arizona Extreme Wavefront Control Lab (XWCL) in partnership with Hart Scientific has manufactured a three-sided pyramid optic with a tip less than $5\mu m$ in size. The optic has been integrated into a 3PWFS that is a visiting instrument on the Comprehensive Adaptive Optics and Coronagraph Test Instrument (CACTI). A future experiment will test the performance of the 3PWFS against the currently integrated 4PWFS under different strengths of turbulence.

\section{Conclusion}

The PWFS is an extension of the Foucault knife edge test to two dimensions. By examining the diffraction theory of the Foucault test, which is much simpler than a full pyramid, we gain insight to the physical processes behind the PWFS. In Section~\ref{diffraction} we consolidated the diffraction theory of the knife edge test by Linfoot\cite{linfoot1948theory}, Katzoff\cite{katzoff1971quantitative}, and Wilson\cite{wilson1975wavefront} into a single derivation with uniform notation. Expanding upon their results, we linked phase aberrations in the shape of Fourier modes to intensity patterns produced by the knife edge test. The result found for a phase error in the shape of a Fourier mode is an intensity pattern described by the Hilbert transform of the phase function. We considered an example $\cos(nx)$ phase pattern, and mathematically derived that the resulting intensity pattern is proportional to $-\sin(nx)$, which is functionally the derivative of the phase without the dependence on spatial frequency. We confirmed this result by considering the intensity and slope signals from a simulated 3PWFS and 4PWFS. We used this result to motivate the signal processing of the pyramid measurements and introduced a new Slope Map method to handling the 3PWFS signals, which is derived using the centroid of an equilateral triangle. This method provided a stable closed-loop correction on the LOOPS testbed.

As a further test of the 3PWFS and slopes maps equations, we performed an end-to-end simulation of an adaptive optics system to compare the performance of the 3PWFS and 4PWFS as well as the Raw Intensity and Slopes Maps signal processing techniques. We found that in the absence of read noise the performance of the wavefront sensors is within 0.01 Strehl. When we included read noise into the simulation, we saw a break in this trend at lower guide star magnitudes and found a gain of 0.036 Strehl the 3PWFS using Raw Intensity over the 4PWFS using Slopes Maps at a stellar magnitude of 10. At the same magnitude, the 4PWFS using Raw Intensity also outperformed the 4PWFS Slopes Maps, but the gain was only 0.0122 Strehl. We conclude that the 3PWFS is a viable wavefront sensor with similar performance in the scope of our study. Our simulations only considered a single modulation radius and $r_0$ of atmospheric turbulence. Further work should explore the respective performances at different turbulence strengths and different modulation radii.

\section*{Acknowledgements}
We thank everyone involved who made the collaboration between the University of Arizona and the Laboratoire d'Astrophysique de Marseille possible. Part of this work was completed during a research visit by Lauren Schatz to the Laboratoire d'Astrophysique de Marseille that was supported by the ARCS Foundation Scholarship. Thank you to Alexander Rodack for his computer help, and Logan Pearce for sharing her Python plotting style sheet. This work has been supported in part by the Air Force Research Laboratory, Directed Energy Directorate, under contract FA9451-19-C-0581. The opinions, findings, and conclusions expressed in this paper are those of the authors and do not necessarily reflect those of the United States Air Force.
This work benefited from the support of the WOLF project ANR-$18$-CE$31$-$0018$ of the French National Research Agency (ANR). It has also been prepared as part of the activities of OPTICON H$2020$ (2017-2020) Work Package $1$ (Calibration and test tools for AO assisted E-ELT instruments). OPTICON is supported by the Horizon 2020 Framework Programme of  the  European  Commission’s  (Grant  number  $730890$). Authors are acknowledging the support by the Action Spécifique Haute Résolution Angulaire (ASHRA) of CNRS/INSU co-funded by CNES. Finally, part of this work is supported by the LabEx FOCUS ANR-$11$-LABX-$0013$, and received the support of Action Spécifique Haute Résolution Angulaire ASHRA.

%%%%% References %%%%%

\bibliography{report}   % bibliography data in report.bib
\bibliographystyle{spiejour}   % makes bibtex use spiejour.bst

%%%%% Biographies of authors %%%%%

\vspace{2ex}\noindent\textbf{Lauren Schatz} is a research physicist at the Air Force Research Laboratory and a recent graduate of the University of Arizona's Wyant College of Optical Science. Her research focuses on developing wavefront sensing instrumentation for high contrast adaptive optics systems. She is a 2018 ARCS Foundation scholar, and a receiver of the Society of Women Engineer's Ada I. Pressman Memorial Scholarship from 2019 to 2021.

\listoffigures
%\listoftables

\end{spacing}
\end{document}